\documentstyle[12pt,aaspp4]{article}
\received{November 7, 1997}
\accepted{February 25, 1997}
\slugcomment{Accepted for publication in the Astrophysical Journal (letters)}
\begin{document}

\title{ROSAT Detection and High Precision Localization of
X-ray Sources in the November 19, 1978 Gamma-Ray Burst Error Box}
\author{M. Bo\"er}
\affil{Centre d'Etude Spatiale des Rayonnements (CNRS/UPS),
BP 4346, F 31028 Toulouse Cedex 4,  France}
\authoremail{boer@cesr.cnes.fr}

\author{C. Motch}
\affil{Observatoire de Strasbourg, 11, rue de l'Universit\'e, F 67000 
Strasbourg, France}

\author{J. Greiner, W. Voges, P. Kahabka}
\affil{Max Planck Institut f\"ur Extraterrestrische Physik, D 85740 Garching, 
Germany}

\author{H. Pedersen}
\affil{NBIfAFG, Copenhagen University Observatory, Juliane Maries Vej 30,
DK-2100 Copenhagen \O, Denmark}

\begin{abstract}

We report on observations of the 1978, November 19
Gamma-Ray Burst source, performed with the ROSAT X-ray HRI experiment.
Two sources were detected, one of which is possibly variable.
The latter source is identical to the source discovered in 1981
by the EINSTEIN satellite, and recently detected by ASCA.
The precise localization of these sources is given, and our data are compared
with optical, radio and previous X-ray data.

\end{abstract}


\section{Introduction}

 Gamma-Ray Bursts (hereafter GRBs) were discovered some 25 years ago, and
their origin remains enigmatic.  Current data and models involve objects at
distances ranging from several hundred of kiloparsecs to cosmological
distances, an uncomfortable ignorance factor of at least $10^6$ (see e.g.
\markcite{La95} Lamb, 1995, Paczy\'nski, 1995\markcite{Pa95}
and following papers for a discussion on the GRB distance
scale). Two approaches have been used to solve this problem,
namely 1) the use of the logN - logS
relation combined with the angular distribution of GRBs, and 2) the search for
counterparts at other wavelengths, either quiescent or transient.

The first X-ray counterpart searches were done with the Einstein
satellite. Five
GRB small error boxes were observed (\markcite{Pi86} Pizzichini et al., 1986).
This resulted in only one marginal, $3.5\sigma$ detection (Grindlay 1982)
\markcite{Gr82} in the case of GRB 781119.  Later, other observations were
 performed with the EXOSAT \markcite{Bo88}\markcite{Bo91}(Bo\"er et al.,
1988, 1991) and ROSAT satellites (\markcite{Bo931}\markcite{Bo932} Bo\"er et
al., 1993a, 1993b, Greiner et al., 1996\markcite{Gr96}), resulting in only one
other possible candidate, that of the GRB 960501 source (\markcite{Hu96a} Hurley
et al. 1996a). In this last case,
the source is seen at low galactic latitude,
is heavily absorbed, and is possibly extragalactic.

The error box of GRB 781119 is one of the smallest known with a size of 8
arcmin$^2$ (Cline et al. 1981 \markcite{Cl81}). It was observed at various
wavelengths and a possible counterpart was found in X-rays in 1981
as mentioned
above. A radio source was found at this position (Hjellming \& Ewald
1981)\markcite{Hj81}. Using archival data, Schaefer (1981)\markcite{Sc81} found 
an 
optical transient object with a position
marginally compatible with the X-ray object,
but its reality was questioned
(\markcite{Zy90}Zytkow 1990;\markcite{Gr90} Greiner et al., 1990; see however
Schaefer 1990\markcite{Sc90}). In addition, two possible emission lines have
been found in the spectrum of GRB 781119 at 420 and 470 keV
(Teegarden \& Cline1980)\markcite{Te80}.

The error box of GRB 781119 was reobserved by Bo\"er et al. (1988) in 1983 in
order to confirm the EINSTEIN detection with the EXOSAT satellite. The source
was not detected, and the $3\sigma$ upper limit on the flux was $1.2 \times
10^{-12} {\rm erg}.{\rm cm}^{-2}.{\rm s}^{-1}$, for a T = $10^6$K 
blackbody 
spectrum. This limit was consistent with
the EINSTEIN detection for a $10^6$K blackbody source at a distance $\geq
2$kpc. On the other hand, the EINSTEIN source, detected at a $3.5\sigma$
level, might have been a variable source, or, alternatively, a
statistical fluctation. In order to settle the question, we observed the 
region
in 1995 with the ROSAT satellite. Because of a technical problem, the
observation was interrupted and rescheduled in 1996. Independently, the same
source was observed by Hurley et al. (1996b)\markcite{Hu96} using the ASCA
satellite.

We report here on the observations carried out by ROSAT,
which are compared with the ASCA observation and data acquired at other
wavelengths. The observation resulted in the detection
of two objects in the error box in 1996, one of them  present in the 1995
observation, possibly variable, and probably associated with the EINSTEIN source.
The other is probably the X-ray counterpart of a quasar.

\section{Observations and Results}

The ROSAT X-ray telescope was used with the HRI detector. The total energy
range is 0.2 - 2.5 keV, with 2 arcsec spatial resolution and no energy
resolution. The data were processed using the EXSAS and MIDAS data analysis
software packages.
The first observation was carried out on January 10,
1995, for a total of 2481 seconds effective time, and the second period of
observations lasted from December 21st, 1995, till January 11th, 1996 for a 
total of
40776 seconds duration time. The observations are summarized
in Table 1.
In the GRB error box, one
object was detected during the 1995 observation, and two objects in 1996.
A SIMBAD cross-identification of the HRI field of view showed that 
two cataloged sources are detected in X-rays: QSO 0116-288, 
and QSO 0117-2837. 

The best position for
object \#1 (detected in both the 1995 and 1996 observations) is
$\alpha = 1{\textstyle h} 18{\textstyle m} 49.6{\textstyle s}$,
$\delta = -28\deg 35' 53"$(equinox 2000.0), and
for object \#2 $\alpha = 1{\textstyle h} 18{\textstyle m} 47.4{\textstyle s}$,
$\delta = -28\deg 35' 45"$, with an error radius of 10 arcsec 
(Briel et al., 1994)
\markcite{Br94}. These positions were computed using the 1996 observation 
only.
As indicated in table 1, there are some marginal indications that 
the flux of object \#1 varied by a factor of
$\approx 2$ between 1995 and 1996, while no evidence for variability was found
during the 1996 observation. 
However, if we look at the flux over an extended period, assuming a constant 
spectrum for the source, long term variability is clearly implied
from the non-detection by the EXOSAT satellite (Bo\"er et al., 1988). 
The non detection of object \#2 in 1995 is
consistent with its flux in 1996. The $3\sigma$ upper limit
to the count rate for object
\#2 in  January 1995 is $4.6 \times 10^{-3}$ c/s.
No variability has been found in the 1996
data for this object.

\section{Discussion}

Figure 1 displays the ROSAT HRI image from the
observation of 1996. We show the EINSTEIN (Grindlay et al., 1982),
ASCA (Hurley et al. 1996b) and ROSAT  error boxes, as well as the radio
sources present in the GRB 781119 error box
(Hjellming and Ewald 1981)\markcite{Hj81}.
There are two HRI sources in the GRB
781119 error box. Object \#2 may be identified with the quasar 
QSO 0116-288
 whose catalog position is only 4 arcsec from the ROSAT position. 
Object \#1 is the closest to the EINSTEIN error box and may be
identified with the radiosource Q of Hjellming and Ewald (1981). This source
is however slighty outside the 90\% confidence EINSTEIN error box, but the low
level of confidence ($3.5\sigma$) of the EINSTEIN detection may introduce some
additional uncertainties in its confidence region. In addition, the proximity
of the HRI 10 arcsec. error box and the EINSTEIN error box is probably
compatible with the possibility that the ROSAT source is in the 99\% EINSTEIN
error box.

More interesting is the fact that the ROSAT source may be
variable. Its flux varied by a factor larger than 2 within roughly one year,
though the uncertainty in the first (January 1995) observation is large.
In addition there is some evidence for variability
between the ASCA observation (Hurley et al. 1996b), and the present data, 
and a clear discrepancy with the EXOSAT observation, which can only
be explained by variability . On
the other hand, there is no evidence for variability in the data taken by
ROSAT during the period December 21st, 1995 - January 11th, 1996. Hence we
deduce that the object \#1 is variable over the long term (i.e. $\geq 1$y).
Because of the lack of spectral resolution
of the ROSAT HRI instrument, we were
unable to investigate whether this variability is also present in the source
spectra.

Thanks to the high precision of the localization given by the ROSAT HRI
instrument, we can reanalyze data taken in 1981 - 1982 at ESO
 (Pedersen et al. 1983)\markcite{Pe83}, as well as more recent,
unpublished observations. The
sum of several unfiltered exposures is displayed in figure 2.
The
uncertainty circles for objects \#1 and \#2 are also displayed. 
A detailed study of the optical data and of
the optical variability of the objects near the ROSAT sources is reported 
by Pedersen (1996)\markcite{Pe96}. From archival data taken by Pedersen et al. 
(1983)
we estimate the R, V, and B magnitudes of object \#2 to be respectively
$\approx 22$, 22.9 , and 23.8.

The probability of having one X-ray source in the GRB error box is 0.16. This 
probability 
has been computed using the local number of sources detected above a 5$\sigma$ 
level within the ROSAT HRI 30'x30' field of view. This probability is consistent 
with the 
ROSAT logN-logS relation derived by Hasinger et al. (1993, 
1994)\markcite{Ha93}\markcite{Ha94} for a 5$\sigma$ detection level of $5 \times 
10^{-14}$erg.cm$^{-2}$.s$^{-1}$ in our observation.
Given the
presence of 2 sources in the error box it is 
even more difficult to associate any of
them with the GRB source. However we note that object \#1 is variable, at least
on a one year time scale, reinforcing the probability of a possible association
with the GRB source. Optical observations are planned at ESO to determine the
precise photometry of the sources, to derive their redshifts, and a possible
association, as they could both belong to the same
cluster of galaxies.

\acknowledgments
HP is grateful for hospitality at NBIfAFG, Copenhagen.
This research has made use of data obtained at the European Southern
Observatory. For data analysis purposes we used the EXSAS package provided by
the Max Planck Institut f\"ur Extraterrestrische Physik, together with the
MIDAS environment from the European Southern Observatory. JG is supported by the 
Deutsche Agentur f\"ur Raumfahrtangelegentheiten (DARA) GmbH under 
contract FKZ 50 OR 9201. 
The ROSAT project is supported by the German Bundes\-mini\-ste\-rium f\"ur 
Bildung und Wissenscchaft (BMBW/DARA) and the Max Planck Society.

%

\clearpage
\begin{deluxetable}{cccc}
\tablecaption{Summary of the ROSAT HRI observations}
\tablehead{
\colhead{Observation date}& \colhead{Jan. 10, 1995}
& \multicolumn{2}{c}{Dec. 21, 1995 - Jan 11, 1996}
}
\startdata
Object \#& 1 & 1 & 2 \nl
Object name& RX J0118.7-2835 & RX J0118.7-2835 & RX J0118.8-2835 \nl
Exposure time (seconds)& 2481 & 40776 & 40776 \nl
Countrate (counts / kseconds) & $6.6 \pm 1.8$ & $2.8 \pm 0.3$ &
$1.2 \pm 0.3$\nl
Signal to noise ratio& 3.6 & 10 & 5.5\nl
Flux\tablenotemark{1} (erg.cm$^{-2}$.s$^{-1}$) &
$4.3 \times 10^{-13}$ & $2.2 \times 10^{-13}$ & $0.8 \times 10^{-13}$ \nl
\tablenotetext{1}{Assuming a power law spectrum of index 1.77 and a hydrogen 
column density of $1.76 \times 10^{20}$ cm$^{-2}$
(Hurley et al. 1996b, Dickey \& Lockman 1990)\markcite{Di90}}
\enddata

\end{deluxetable}


\clearpage

\begin{figure}
\caption{A 15 x 15 arcmin close-up of the ROSAT HRI image centered on
the GRB 781119 error box, taken
during the period December 21st, 1995 - January 11th, 1996. The error box of
the GRB source is displayed (polygon), as well as the Einstein (small circle) 
and ASCA 
(large circle) error regions, the HRI sources
\#1 and \#2, the radio objects B, C, and Q detected by Hjellming and Ewald 
(1981), and 
the position of the quasar source QSO 0116-288 (cross).}
\label{fig1}
\end{figure}

\begin{figure}
\caption{The optical content of the region the ROSAT HRI source \#1 and \#2.
The image is the sum of several unfiltered CCD exposures from ESO
}
\label{fig2}
\end{figure}

\begin{figure}
\plotone{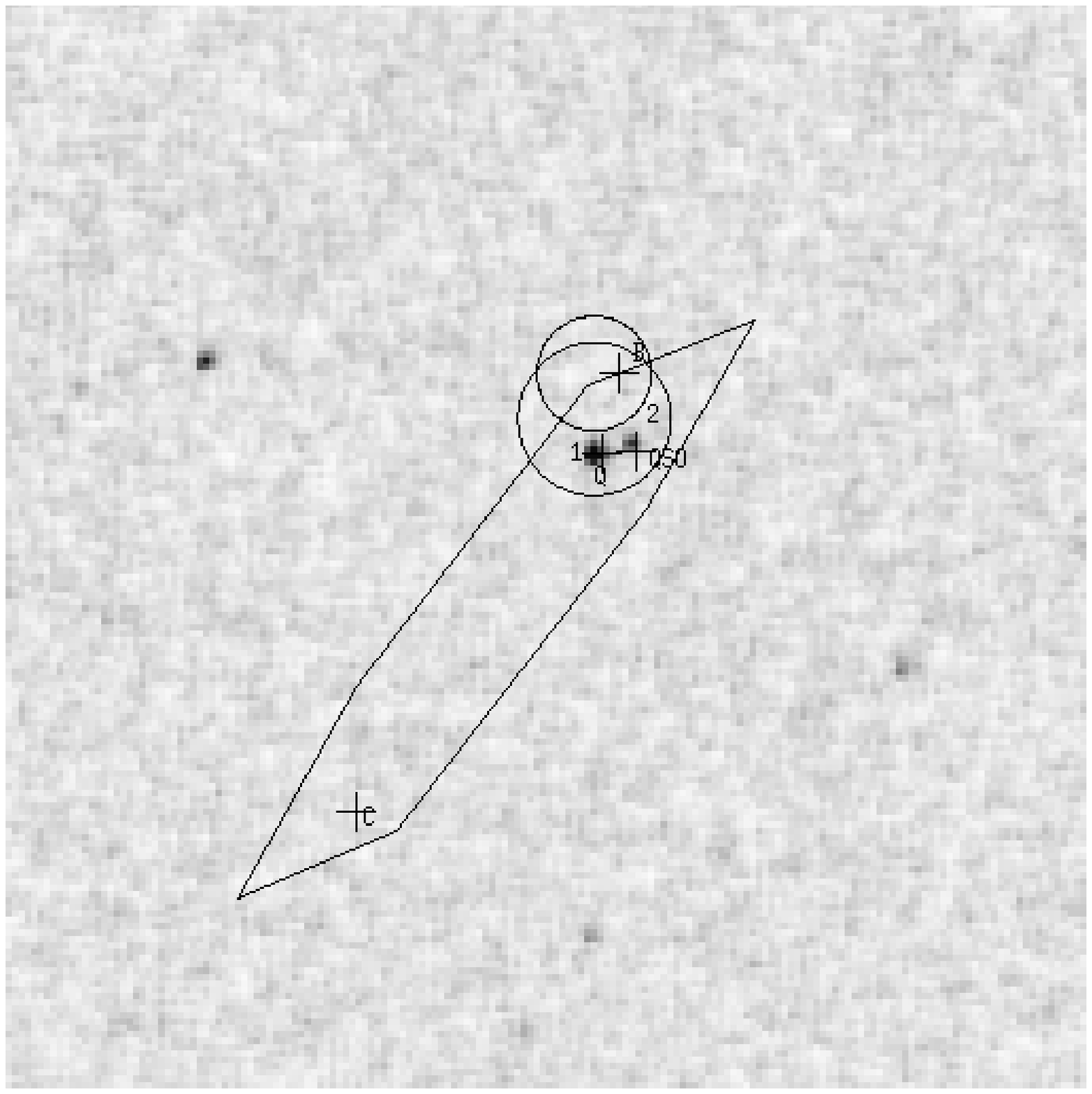}
\end{figure}

\begin{figure}
\plotone{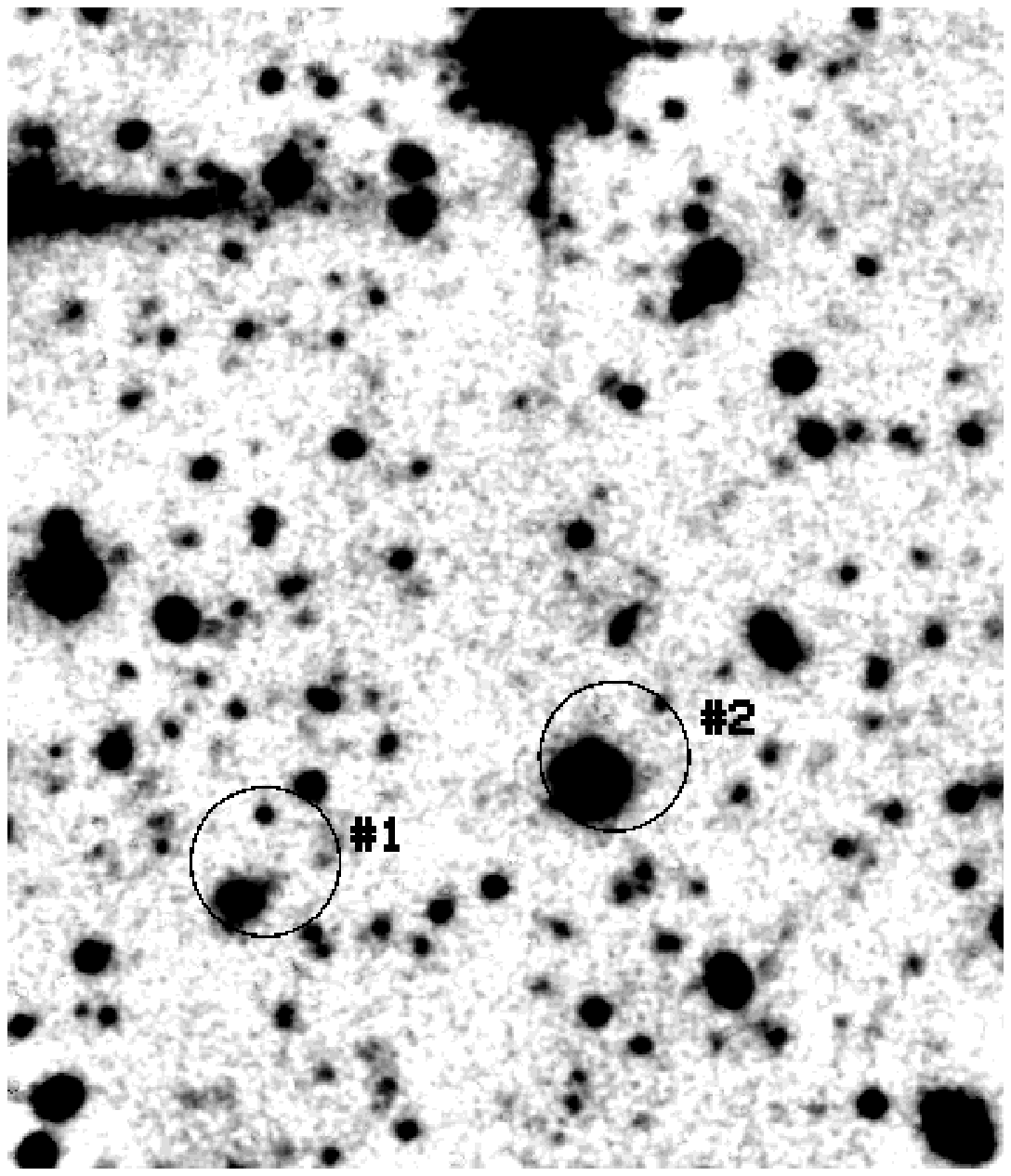}
\end{figure}
\end{document}